\documentclass[%
 reprint,
superscriptaddress,
 amsmath,amssymb,
 aps,
]{revtex4-2}

\usepackage{graphicx}
\usepackage{dcolumn}
\usepackage{bm}
\usepackage{multirow}
\usepackage{color}
\usepackage{hyperref}
\definecolor{darkblue}{rgb}{0,0,0.5}
\hypersetup{
colorlinks=true,
linkcolor=black,
filecolor=blue,
citecolor=darkblue,  
urlcolor=black,
}

\newcommand{\GK}[1]{{{\textcolor{black}{#1}}}}

\begin{document}

\preprint{APS/123-QED}

\title{Mixed-State Quantum Denoising Diffusion Probabilistic Model}

\author{Gino Kwun}
\affiliation{Thomas Lord Department of Computer Science, University of Southern California, Los Angeles, California 90089, USA}

\author{Bingzhi Zhang}
\email{bingzhiz@usc.edu}
\affiliation{ Department of Physics and Astronomy, University of Southern California, Los
Angeles, California 90089, USA
}
\affiliation{
Ming Hsieh Department of Electrical and Computer Engineering, University of Southern California, Los
Angeles, California 90089, USA
}
\author{Quntao Zhuang}
\email{qzhuang@usc.edu}
\affiliation{
Ming Hsieh Department of Electrical and Computer Engineering, University of Southern California, Los
Angeles, California 90089, USA
}
\affiliation{ Department of Physics and Astronomy, University of Southern California, Los
Angeles, California 90089, USA
}

\date{\today}

\begin{abstract}
Generative quantum machine learning has gained significant attention for its ability to produce quantum states with desired distributions. Among various quantum generative models, quantum denoising diffusion probabilistic models (QuDDPMs) [Phys. Rev. Lett. 132, 100602 (2024)] provide a promising approach with stepwise learning that resolves the training issues. However, the requirement of high-fidelity scrambling unitaries in QuDDPM poses a challenge in near-term implementation. We propose the \textit{mixed-state quantum denoising diffusion probabilistic model} (MSQuDDPM) to eliminate the need for scrambling unitaries. Our approach focuses on adapting the quantum noise channels to the model architecture, which integrates depolarizing noise channels in the forward diffusion process and parameterized quantum circuits with projective measurements in the backward denoising steps. We also introduce several techniques to improve MSQuDDPM, including a cosine-exponent schedule of noise interpolation, the use of single-qubit random ancilla, and superfidelity-based cost functions to enhance the convergence. We evaluate MSQuDDPM on quantum ensemble generation tasks, demonstrating its successful performance. 


\end{abstract}

\maketitle


\section{\label{sec:level1}Introduction}

Quantum machine learning (QML), integrating quantum mechanical principles such as entanglement and superposition to optimize and approximate complex functional behavior, has demonstrated superior performance compared to classical approaches in certain optimization problems and potential efficiency in specific contexts \cite{biamonte2017quantum, farhi2014qaoa, Amin_2018QBM, rebentrost2014qsvm}. Variational quantum algorithms (VQAs), designed using parameterized quantum circuits (PQCs) or variational quantum circuits (VQCs) with classical optimization, have enabled the development of QML algorithms in conjunction with various applications on noisy intermediate-scale quantum (NISQ) devices  \cite{cerezo2021vqa, woerner2019quantumriskanalysis, maheshwari2022quantumbio}. 

Classical deep generative models (DGMs) such as generative adversarial network (GAN) \cite{goodfellow2020generative}, variational autoencoder (VAE) \cite{kingma2013auto}, and attention-based models, including Transformers \cite{vaswani2017attention}, have substantially advanced machine learning by producing realistic samples across various domains such as vision \cite{wang2021gancv}, language \cite{openai2023gpt}, speech \cite{tan2021surveyneuralspeechsynthesis}, and audio generation \cite{dhariwal2020jukeboxgenerativemodelmusic}. Among the plethora of versatile generative models, denoising diffusion probabilistic models (DDPMs) and their variations \cite{ho2020denoisingdiffusionprobabilisticmodels, zhang2022gddim, kingma2023variationaldiffusionmodels}, which represent a branch of hierarchical VAEs incorporating a thermodynamic approach in deep learning \cite{sohl2015deep, yang2023diffusion}, have provided another way to create realistic samples. The model includes two main components. The forward process gradually injects Gaussian noise into the data and transforms it into complete Gaussian white noise. 
The backward process learns from intermediate forward samples to effectively remove noise using deep neural network. 


Inspired by the success of classical DGMs, quantum generative models, including quantum generative adversarial network (QuGAN) \cite{lloyd2018qugan, zoufal2019quantum, huang2021experimental}, quantum variational autoencoder (QVAE) \cite{khoshaman2018quantum}, tensor network \cite{wall2021generative}, and diffusion-based quantum generative model \cite{BingzhiQuDDPM, cacioppo2023quantum, parigi2024quantum, chen2024quantum, zhu2024quantum} have been presented and studied extensively for their potential in quantum data generation and their applications in real-world tasks. In particular, the quantum denoising diffusion probabilistic models (QuDDPMs)~\cite{BingzhiQuDDPM} adopt quantum scrambling as the forward process and VQCs with projective measurements as the backward process. Via the forward scrambling process creating interpolation between target and noise, QuDDPM resolves the training issue that plagues VQC-based algorithms~\cite{mcclean2018barren, Cerezo_2021}. However, the scrambling-based forward process requires multiple steps of high-fidelity unitaries, creating a challenge in near-term implementation. In addition, the implementations of QuDDPM in Ref.~\cite{BingzhiQuDDPM} are restricted to pure states, leaving the general mixed state unexplored.

To address the limitations, we propose the \textit{mixed-state quantum denoising diffusion probabilistic model} (MSQuDDPM) as a generalization of the QuDDPM framework. This model adopts the depolarizing noise channels \cite{king2002capacityquantumdepolarizingchannel} as the forward process and maintains PQCs with projective measurements as the backward process \cite{BingzhiQuDDPM}. MSQuDDPM extends the capability of QuDDPMs to produce both pure and mixed quantum state ensembles and provides computationally efficient training and sample generation. Compared to the original QuDDPM designed for pure state generation, the protocol in this work achieves comparable performance in the same tasks while reducing the implementation complexity. By utilizing superfidelity-based cost functions, adapting single-qubit random states as an ancilla in model training, and various noise scheduling approaches, we further enhance the performance of MSQuDDPM. MSQuDDPM extends the utility of quantum generative models and narrows the gap toward practical utilization. 

The paper is organized as follows. We describe the design of the model architecture and training strategy of MSQuDDPM in Section~\ref{sec:formulation}. We then present and analyze the results of applying MSQuDDPM to several quantum state ensemble generation tasks in Section~\ref{sec:applications}. Section~\ref{sec:improveStretegy} discusses the strategies to enhance MSQuDDPM's capability through dividing diffusion steps, noise scheduling in multi-qubit ensemble generation, and ancillary qubit initialization. We finally conclude our work and briefly discuss potential directions for future research.



\section{\label{sec:formulation}Formulation of MSQuDDPM}

\begin{figure}
\includegraphics[width=0.48\textwidth]{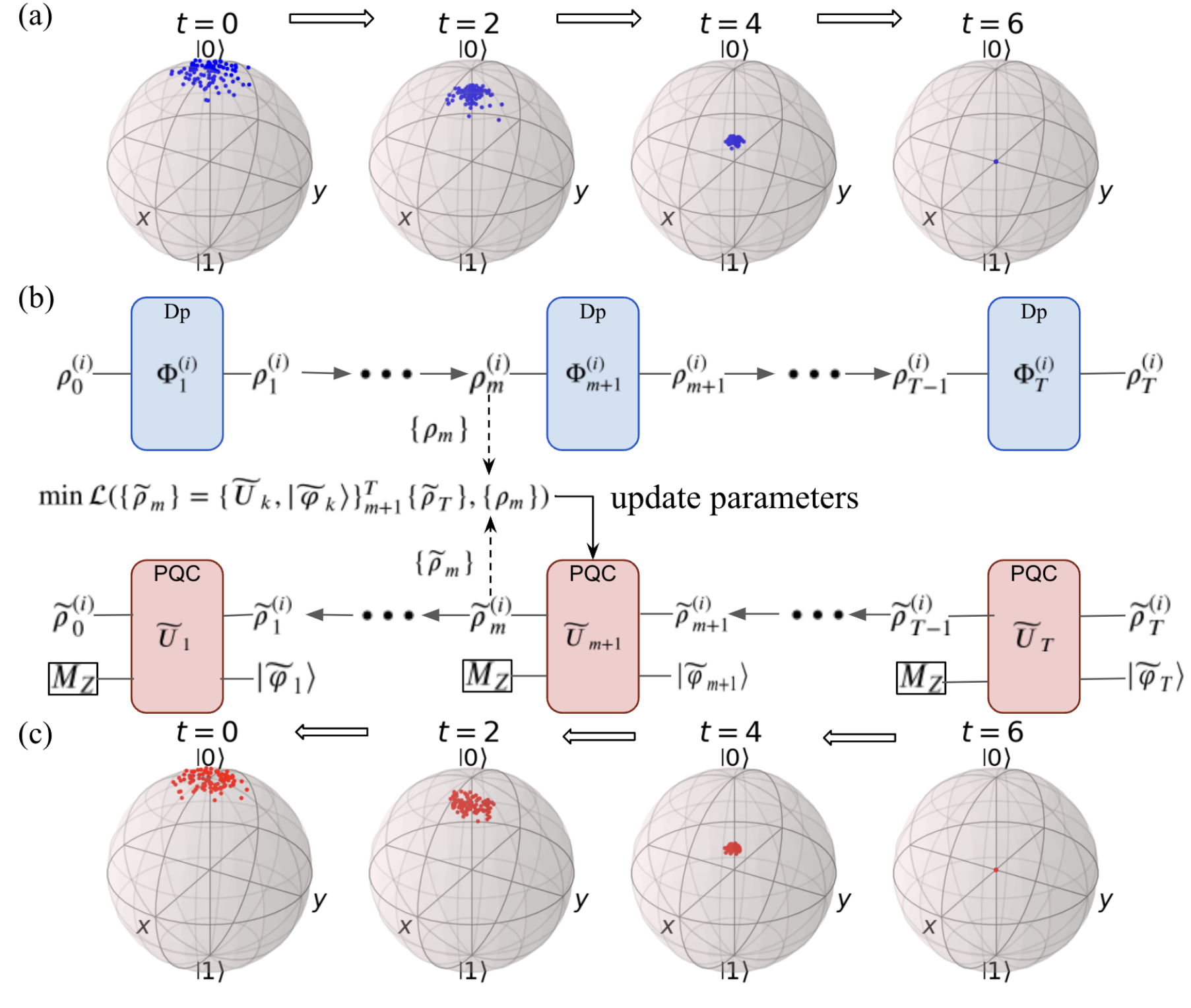}
\caption{\label{fig:modelarch}The model architecture and training strategy of MSQuDDPM are described in (b). The forward process utilizes depolarizing channels to inject noise, while the backward process in the inverse order is constructed with PQCs to learn the noise removal. (a) and (c) show MSQuDDPM's forward process samples from $t=0$ to $t=6$ and its backward procedure samples from $t = 6$ to $t=0$ in a single-qubit learning task. See Table~\ref{tab:expresult} for details.}
\end{figure}

In this section, we describe the architecture of MSQuDDPM, focusing on its training strategy and cost functions. MSQuDDPM contains two main components: a forward diffusion process and a backward denoising process, as illustrated in Fig.~\ref{fig:modelarch} (b). Similar to classical DDPMs, MSQuDDPM gradually injects noise to convert the original distribution at $t = 0$ into simple but noisy distributions, such as Haar random states and maximally mixed states at $t=T$ \cite{ho2020denoisingdiffusionprobabilisticmodels, BingzhiQuDDPM}. The backward process uses PQCs to sequentially approximate the recovery of the noise induced by the forward diffusion from $t = T$ to $t = 0$. 

\subsection{\label{sec:level2}Forward diffusion process}

The forward diffusion process begins with an $n$-qubit initial quantum state ensemble $\{\rho_0\}$, consisting of $n_\text{data}$ individual quantum states (pure or mixed), sampled from an unknown distribution. Throughout the process, the model iteratively applies $T$ discrete quantum channels $\Phi_1, \Phi_2, \cdots, \Phi_{T}$, generating incremental noisy ensembles $\{\rho_1\}, \{\rho_2\}, \cdots, \{\rho_T\}$. We employ a depolarizing channel as a noise provider, which leads the state ensemble toward a totally mixed state \cite{Nielsen_Chuang_2010, king2002capacityquantumdepolarizingchannel}. In detail, at time $t$, the relation between the states $\rho_t^{(i)} \in \{\rho_t\}$ and $\rho_{t+1}^{(i)} \in \{\rho_{t+1}\}$ is given by 
\begin{equation}
    \rho_{t+1}^{(i)} = \Phi_{t+1} (\rho_t^{(i)}) = (1-q_{t+1}^{(i)})\rho_t^{(i)} + q_{t+1}^{(i)} I / d,
\end{equation}
where $q_{t+1}^{(i)} \in (0, 1]$ is the depolarizing parameter, and $I/d$  represents the maximally mixed state.

For the noise scheduling of depolarizing parameters, we adopt linear \cite{ho2020denoisingdiffusionprobabilisticmodels} and cosine scheduling \cite{nichol2021improveddenoisingdiffusionprobabilistic} from classical DDPM. However, in the multi-qubit system, the forward samples rapidly converge to maximally mixed states, precluding meaningful training during the early stages of the backward process. Thus, we propose the \textit{cosine-exponent schedule} to amplify the preservation effect of the initial state ensemble, denoted by 

\begin{eqnarray}
q_t = (1 - \frac{\overline{\alpha}_t}{\overline{\alpha}_{t-1}})^k
\label{eq:cosineExponent},
\end{eqnarray}
where
\begin{eqnarray}
\overline{\alpha}_t = \frac{f(t)}{f(0)}, f(t) = \cos{(\frac{t/T + \epsilon}{1 + \epsilon}\cdot \pi/2)^2}
\label{eq:six},
\end{eqnarray}
$k$ is a constant, and $\epsilon$ is a small offset. In this work, we use cosine and cosine square scheduling, corresponding to $k=1$ and $k=2$, respectively. Generation of the final state $\rho_T^{(i)} \in \{\rho_T\}$ can be achieved by repetitively applying the channels, i.e., $\rho_T^{(i)} = \Phi_{T}\circ \Phi_{T-1}\circ \cdots \circ \Phi_2\circ \Phi_1(\rho_0^{(i)})$. 

\subsection{\label{sec:backwardProcess}Backward denoising process}

The backward denoising process begins with an ensemble $\{\tilde{\rho}_T\}$ of $n_\text{train}$ copies of $n$-qubit maximally mixed states and gradually produces $\{\tilde{\rho}_0\}$ that resembles the original state ensemble $\{\rho_0\}$. This process utilizes a VQA with $T$ total chunks of sequentially trainable VQCs and projective measurements with $n_a$ ancillary qubits \cite{BingzhiQuDDPM}. We considered two initialization methods for the ancillary qubits, one uses all zero states, ($|0\rangle ^{\otimes n_a}$), and the other replaces a single qubit with a Haar random pure state, ($|\phi_{\rm Haar} \rangle \otimes (|0\rangle)^{\otimes (n_a - 1)}$), to introduce randomness and therefore enhance the generative power. We sample the $n_\text{train}$ Haar random states for each diffusion step and maintain these ancillary states throughout the individual training process. After training, we freshly draw the states for sampling and input generation in subsequent stages. The final quantum state ensemble $\{\tilde{\rho}_0\}$ is produced by running the sequence of trained PQCs in reverse order, using $n_{\rm test}$ samples of maximally mixed states.
\GK{The circuit layout of the forward and backward process is presented in Appendix~\ref{sec:circuitLayout}. }

Having the forward and backward quantum circuits, Fig.~\ref{fig:modelarch}(b) illustrates how training proceeds at stage $t=m+1$. The forward depolarizing model generates $n_{\text{train}}$ samples in the target ensemble $\{\rho_m\}$ with $\Phi_{1}, \Phi_{2}, \cdots,\Phi_{m-1}$, and $ \Phi_m$, while the backward PQC outputs the ensemble $\{\tilde{\rho}_m\}$ using $\tilde{U}_{T},  \tilde{U}_{T-1}, \cdots, \tilde{U}_{m+2}$, and $ \tilde{U}_{m+1}$. Based on the loss function that captures the resemblance between the output ensemble $\{\tilde{\rho}_{m}\}$ and the forward depolarizing result $\{\rho_{m}\}$, the gradient-based algorithm trains the parameters $\bm \theta_{m+1}$ of the PQC. After optimization, the parameters $\bm \theta_{m+1}$ are fixed and reused in further steps. 
\GK{By leveraging this `divide-and-conquer' strategy, MSQuDDPM allows easier optimization at each stage with relatively small discrepancies between consecutive ensembles. Following the QuDDPM's trainability analysis \cite{BingzhiQuDDPM}, suppose the model requires at least a polynomial circuit depth in the system size $n$ to generate quantum ensembles. By splitting the problem into multiple subproblems with circuit depth scaling logarithmically with the system size, $\log(n)$, MsQuDDPM can maintain shallow circuits in each step to ensure convergence and avoid training issues such as the barren plateau \cite{mcclean2018barren, Cerezo_2021}.}

\subsection{\label{sec:costFun}Cost function}

The cost function of MSQuDDPM is based on the similarity between two quantum state ensembles. For individual states, this similarity can be quantified using fidelity \cite{Jozsa1994FidelityFM}. Fidelity $F(\rho, \sigma)$, measuring the similarity between two $n$-qubit density matrices $\rho$ and $\sigma$, is defined as $F(\rho, \sigma) = [\text{Tr}(\sqrt{\sqrt{\rho}\sigma\sqrt{\rho}})]^2$,
where $\text{Tr}$ denotes the trace operation. However, fidelity-based cost functions cannot be effective in MSQuDDPM because they may require full tomography of mixed states, which would require a large number of copies of the state, making it impractical in high-dimensional systems \cite{schmale2022qst}. To address the limitations of fidelity, we use \textit{superfidelity}, an upper-bound approximation of fidelity, due to its computational feasibility in mixed states, requiring fewer state copies and alleviating potential inefficiencies \cite{miszczak2008subsuperfidelityboundsquantum}. The superfidelity between two density matrices $\rho$ and $\sigma$ is computed by 

\begin{eqnarray}
G(\rho, \sigma) = \text{Tr}(\rho\sigma) + \sqrt{[1 - \text{Tr}(\rho^2)][1-\text{Tr}(\sigma^2)]}
\label{eq:two}.
\end{eqnarray}

Extending this notion to quantum state ensembles, we define the \textit{mean superfidelity} $\overline{G}(\{\rho_1\}, \{\rho_2\})$ of two quantum state ensembles $\{\rho_1\}$ and $\{\rho_2\}$ as

\begin{eqnarray}
\overline{G}(\{\rho_1\}, \{\rho_2\}) = \mathbb{E}_{\rho \sim \{\rho_1\}, \sigma \sim \{\rho_2\}}[G(\rho, \sigma)],
\label{eq:three}
\end{eqnarray}
and leverage this metric to measure the discrepancy between two quantum state ensembles. Here, we assume that $\rho$ and $\sigma$ are randomly sampled from ensembles $\{\rho_1\}$ and $\{\rho_2\}$, respectively. 

The loss functions are extensions of Ref.~\cite{BingzhiQuDDPM} to mixed states, which can be maximum mean discrepancy (MMD) \cite{gretton12a} and Wasserstein distance \cite{Ognyan2009Wasserstein} with superfidelity as a kernel function and the distance measure, respectively. In the learning cycle at $t=m+1$ in Fig.~\ref{fig:modelarch} (a), the model constructs two ensembles, one from the backward model outputs $\{\tilde{\rho}_{m}\}$ and the other from the target states $\{\rho_m\}$. Then, the training algorithm minimizes the (squared) MMD distance between the two quantum state ensembles, defined as

\begin{eqnarray}
D_{\rm MMD}(\{\tilde{\rho}_m\}, \{\rho_m\}) = &&\overline{G}(\{\tilde{\rho}_m\}, \{\tilde{\rho}_m\}) + \overline{G}(\{\rho_m\}, \{\rho_m\})\nonumber\\
&&- 2\overline{G}(\{\tilde{\rho}_m\}, \{\rho_m\})
\label{eq:four}.
\end{eqnarray}
The Wasserstein distance is presented as an enhancement for generating complex quantum state ensembles where it is infeasible to distinguish between two state ensembles with MMD distance~\cite{BingzhiQuDDPM}. After reducing Kantorovich's formulation of the optimal transport problem into linear programming through discretizing and limiting the mass distributions of quantum state ensembles $\{\rho_i\}_{i=1}^{v}$ and $\{\sigma_j\}_{j=1}^w$ with sizes $v$ and $w$, respectively, we provide the transport plan cost as $v \times w$ pairwise superfidelity-based cost measure $C_{i, j} = 1 - G(\rho_i, \sigma_j)$ between two quantum states from each ensemble. The minimum total distance of the corresponding optimal transport problem is expressed as
\begin{eqnarray}
\text{Wass}(\{\rho\}, \{\sigma\}) =\min_{P} &&\langle P, C \rangle,\\
\text{s.t.} &&P{\bm 1}_w = r,\nonumber\\
&& P^{\intercal}{\bm 1}_v = s,\nonumber\\
&& P \geq 0,\nonumber
\end{eqnarray}
where $\langle \cdot, \cdot\rangle$ denotes the inner product, $P$ is the transport plan, $\bm 1_v$ and $\bm 1_w$ are all-ones vectors, and $r$ and $s$ are the probability vectors corresponding to $\{\rho\}$ and $\{\sigma\}$, respectively.

\section{\label{sec:applications}Applications}

\begin{table*}
\begin{ruledtabular}
\begin{tabular}{c||c|c|c|c|c|c|c|c|c}
 Task&$n$&$n_{a}$&$n_{\rm train}$
&$T$&$L$&Cost function&Forward schedule&Ancilla type &Performance\\
\hline
\multirow{2}{10em}{Clustered states (Fig.~\ref{fig:modelarch} and Fig.~\ref{fig:HaarComp} a2)}&\multirow{2}{*}{1}&\multirow{2}{*}{2}&\multirow{2}{*}{100}&\multirow{2}{*}{6}&\multirow{2}{*}{4}&\multirow{2}{*}{Wasserstein}&\multirow{2}{*}{Cosine}&\multirow{2}{*}{Haar ($|\phi_{\rm Haar}\rangle \otimes |0\rangle$)} &$\overline{F}_{\text{data}, 0} = 0.9853 \pm 0.0001$\\
 & & & & & & & & &$\overline{F}_{\text{gen}, 0} = 0.9873 \pm 10^{-5}$\\\hline
\multirow{2}{8em}{Clustered states (Fig.~\ref{fig:HaarComp} a3)}&\multirow{2}{*}{1}&\multirow{2}{*}{2}&\multirow{2}{*}{100}&\multirow{2}{*}{6}&\multirow{2}{*}{4}&\multirow{2}{*}{Wasserstein}&\multirow{2}{*}{Cosine}&\multirow{2}{*}{zero ($|00\rangle$)} &$\overline{F}_{\text{data}, 0} = 0.9852\pm 0.0001$\\
 & & & & & & & & &$\overline{F}_{\text{gen}, 0} = 0.9888 \pm 10^{-5}$\\\hline
\multirow{2}{10em}{Circular states (Fig.~\ref{fig:circular} and Fig.~\ref{fig:HaarComp} b3)}&\multirow{2}{*}{1}&\multirow{2}{*}{2}&\multirow{2}{*}{200}&\multirow{2}{*}{6}&\multirow{2}{*}{8}&\multirow{2}{*}{Wasserstein}&\multirow{2}{*}{Cosine square}&\multirow{2}{*}{zero ($|00\rangle$)} &${\rm Wass}_{\rm data} = 0.0063$\\
 & & & & & & & & &${\rm Wass }_{\text{gen}} = 0.0151$\\\hline
\multirow{2}{8em}{Circular states (Fig.~\ref{fig:HaarComp} b2)}&\multirow{2}{*}{1}&\multirow{2}{*}{2}&\multirow{2}{*}{200}&\multirow{2}{*}{4}&\multirow{2}{*}{8}&\multirow{2}{*}{Wasserstein}&\multirow{2}{*}{Cosine square}&\multirow{2}{*}{Haar ($|\phi_{\rm Haar}\rangle \otimes |0\rangle$)} &${\rm Wass}_{\rm data} = 0.0068$\\
 & & & & & & & & &${\rm Wass }_{\text{gen}} = 0.0163$\\\hline
\multirow{2}{8em}{Many-body phase (Fig.~\ref{fig:mbpRes} and Fig.~\ref{fig:forwardComp})}&\multirow{2}{*}{4}&\multirow{2}{*}{2}&\multirow{2}{*}{100}&\multirow{2}{*}{6}&\multirow{2}{*}{12}&\multirow{2}{*}{MMD}&\multirow{2}{*}{Cosine square}&\multirow{2}{*}{zero ($|00\rangle$)} &$\overline{M}_{x, \text{data}} = 0.951 \pm 3 \times 10^{-5}$\\
& & & & & & & & &$\overline{M}_{x, \text{gen}} = 0.940 \pm 0.0005$\\\hline
\multirow{2}{8em}{Many-body phase (Fig.~\ref{fig:forwardComp})}&\multirow{2}{*}{4}&\multirow{2}{*}{2}&\multirow{2}{*}{100}&\multirow{2}{*}{6}&\multirow{2}{*}{12}&\multirow{2}{*}{MMD}&\multirow{2}{*}{Linear}&\multirow{2}{*}{zero ($|00\rangle$)} &$\overline{M}_{x, \text{data}} = 0.952 \pm 3\times10^{-5}$\\
& & & & & & & & &$\overline{M}_{x, \text{gen}} = 0.427 \pm 0.0343$\\\hline
\multirow{2}{8em}{Many-body phase (Fig.~\ref{fig:forwardComp})}&\multirow{2}{*}{4}&\multirow{2}{*}{2}&\multirow{2}{*}{100}&\multirow{2}{*}{6}&\multirow{2}{*}{12}&\multirow{2}{*}{MMD}&\multirow{2}{*}{Cosine}&\multirow{2}{*}{zero ($|00\rangle$)} &$\overline{M}_{x, \text{data}} = 0.952 \pm 3 \times 10^{-5}$\\
& & & & & & & & &$\overline{M}_{x, \text{gen}} = 0.679 \pm 0.0232$\\

\end{tabular}
\caption{\label{tab:expresult}Details of the hyperparameters and results for MSQuDDPM application tasks. To assess performance after training, the trained PQC is applied to $n_{\rm test}$ samples of maximally mixed states and the designated metric is compared with $n_{\rm test}$ samples from the original quantum state ensemble. For the clustered states task, we used the mean fidelity between the model output ensemble and $|0\rangle$ defined as $\overline{F}_0 (\{\rho_t\}) = \sum_{\rho \in \{\rho_t\}} \langle 0|\rho|0\rangle/|\{\rho_t\}|$,  where $|\{\rho_t\}|$ indicates the size of the ensemble $\{\rho_t\}$. The Wasserstein distance is computed to measure the resemblance between the data $\{\rho_0'\}$ sampled from the goal distribution and backward test samples in circular states. In the many-body phase learning task, we quantify the average X-axis magnetization defined as $\overline{M}_x (\{\rho_t\}) = \sum_{\rho \in \{\rho_t\}} {\rm Tr} (\rho \sum_i X_i)/(n|\{\rho_t\}|)$. Here, $\overline{F}_{\rm data, 0}$ indicates the mean fidelity between the initial quantum state ensemble and $|0\rangle$, while $\overline{F}_{\rm gen, 0}$ represents the mean fidelity between the generated test ensemble and $|0\rangle$. }
\end{ruledtabular}
\end{table*}

There are several tasks utilizing MSQuDDPM to generate the desired quantum state ensembles. In this section, we present these tasks, their results, and interpretations. Since the initial data for training and testing of the backward process are all maximally mixed states, we use the forward diffusion and backward testing samples for visualization. 
\GK{Note that the initial parameters of each task are set as examples to evaluate MSQuDDPM’s performance and can be adjusted to accomodate various problem setups.}
\GK{Details of the simulations and implementation are provided in Table~\ref{tab:expresult} and Appendix~\ref{sec:implDetail}, respectively.}

\subsection{\label{sec:cluster}Clustered states}

In the clustered state generation task, the initial state ensemble is the set of density matrices $\{\rho_0\}$, where each $\rho_0$ is defined as $\rho_0 =(1-q_0 )|\psi_0 \rangle \langle \psi_0 | + q_0 I / d$. 
\GK{Here, $q_0 \in [0, 0.01)$ is the depolarizing parameter to construct the initial ensemble with mixed states, $I/d$ represents the maximally mixed state, and $|\psi_0 \rangle = |0\rangle + \epsilon_0 c_0 |1\rangle$ up to normalization coefficient with $\epsilon_0 = 0.08$, and ${\rm Re}(c_0), {\rm Im}(c_0) \sim \mathcal{N}(0, 1)$.}
We employ two ancillary qubits, initialized as $|\phi_{\rm Haar}\rangle \otimes |0\rangle$. The training of the backward PQC uses the Wasserstein distance cost function with superfidelity between the backward circuit outputs at step $t$, $\{\tilde{\rho}_{t-1}\}$, and forward diffusion results at step $t-1$, $\{\rho_{t-1}\}$. Fig.~\ref{fig:modelarch}(a) and (c) depict samples from the forward and backward processes, illustrating MSQuDDPM's ability to recover the diffusion process of the clustered distribution.

\subsection{\label{sec:circular}Circular states}

\begin{figure}
\includegraphics[width=0.45\textwidth]{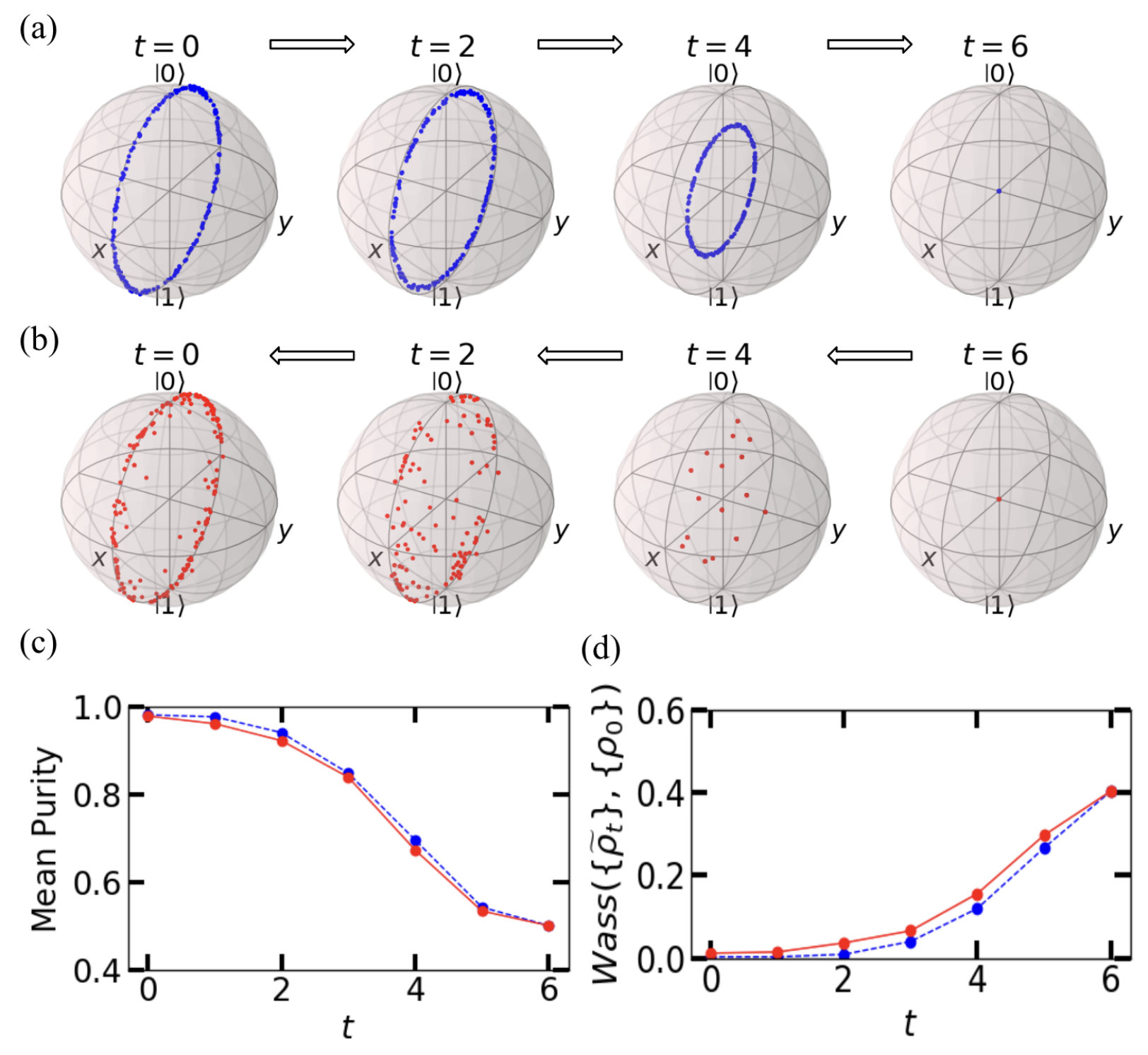}
\caption{\label{fig:circular} Bloch sphere representations of (a) forward and (b) backward testing samples for the circular state ensemble generation. (c) displays the mean purity decay for the forward (blue) and the backward (red) results. The Wasserstein distance decay, $\text{Wass}(\{\tilde{\rho}_t\}, \{\rho_0\})$, for the forward (blue) and the backward (red) outcomes is plotted in (d). The Wasserstein distance is computed between the model outputs $\{\rho_t\}$ or $\{\tilde{\rho}_t\}$ and the initial input samples $\{\rho_0\}$. }
\end{figure}

The initial state ensemble for the circular state case forms a depolarized circle in the X-Z plane of the Bloch sphere, defined by the single-qubit density matrices $\{\rho_0\}$, where each $\rho_0$ is represented as $\rho_0 = (1-q_0 )|\psi_0 \rangle \langle \psi_0 | + q_0 I / d$. The depolarizing parameter $q_0$ is sampled from the interval $[0, 0.04)$, and each $|\psi_0\rangle$ is initialized as $|\psi_0\rangle = RY_{\theta_0}(|0\rangle)$, where $\theta_0 \in [0, 2\pi)$ indicates a rotation around the Y-axis.

In the backward process, we utilize two ancillary qubits in the state $|00\rangle$ and employ the Wasserstein distance with superfidelity to reconstruct quantum states in a ring formation. As depicted in Fig.~\ref{fig:circular}(a) and (b), the generated samples resemble the forward samples as the backward denoising proceeds. Furthermore, Fig.~\ref{fig:circular}(c) and (d) show the decay of the average purity and the Wasserstein distance, maintaining similar trends between the depolarizing and reverse processes.

\subsection{\label{sec:mbp}Many-body phase learning}

\begin{figure}
\includegraphics[width=0.39\textwidth]{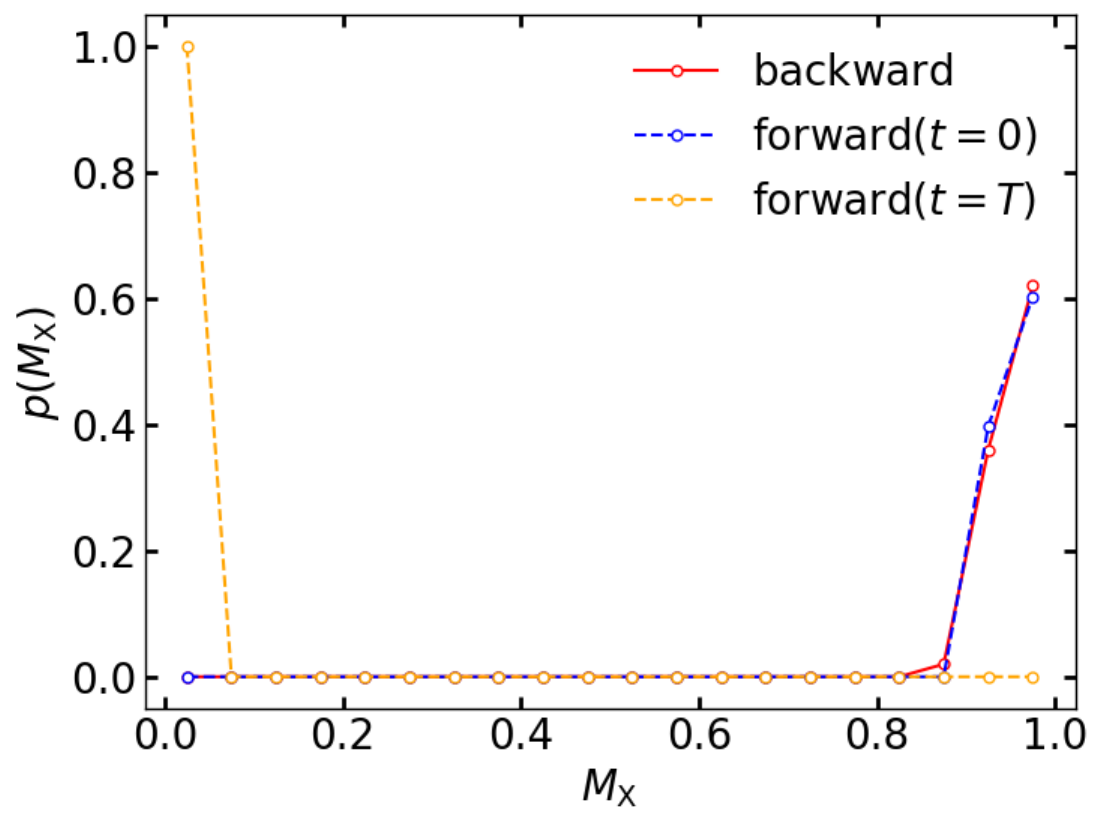}
\caption{\label{fig:mbpRes} Distribution of magnetization along the X-axis ($M_x$) for the forward and backward results in the many-body phase learning task. The red line represents the distribution of backward test samples at $t = 0$. The blue and orange lines denote the X-axis magnetization distribution of the forward initial ensemble ($t = 0$) and the forward final ensemble ($t = T$), respectively. }
\end{figure}

To test our model under more complex conditions, we applied MSQuDDPM to the transverse-field Ising model (TFIM) in condensed matter physics \cite{pfeuty1970one}. The Hamiltonian of TFIM is given by: $H_{\text{TFIM}} = -(\sum_i Z_i Z_{i+1} + g \sum_i X_i)$, where $Z_i$ and $X_i$ denote the Pauli-Z and Pauli-X operations acting only on qubit $i$. As the parameter $g$ increases from zero, a phase transition occurs in the system from an ordered phase ($|g| < 1$) to a disordered phase ($|g| > 1$), with the critical point at $g = 1$. We consider the 4-qubit case ($n=4$) and sample initial states from the ground states of $H_{\text{TFIM}}$ in the disordered phase, which are paramagnetic along the X-axis, where $g$ is uniformly sampled from the range $[1.8, 2.2)$. To evaluate the model's performance, we calculate the X-axis magnetization, $M_x (\rho) = {\rm Tr} (\rho \sum_i X_i)/n$, of the generated samples to determine the phase of the density matrix. As shown in Fig.~\ref{fig:mbpRes}, the majority of the generated states can recover their original magnetization, in contrast to the magnetization of totally mixed states. 

\section{\label{sec:improveStretegy} Strategies for improving MSQuDDPM}

We propose several suggestions that we have gained from MSQuDDPM simulations for better model utilization. Our findings focus on three aspects of MSQuDDPM. First, we address the design of MSQuDDPM and compare the effects of splitting diffusion steps with increasing the number of ancillary qubits for generating high-quality samples and robust scaling. Moreover, we consider the noise scheduling in the forward process of MSQuDDPM for multi-qubit ensemble production. Finally, we investigate the advantage of using single-qubit Haar random states as an auxiliary qubit to diversify the generated samples.

\subsection{\label{sec:designPrinciple} Design principle: more diffusion steps or more ancillary qubits}

\begin{figure}
\includegraphics[width=0.4\textwidth]{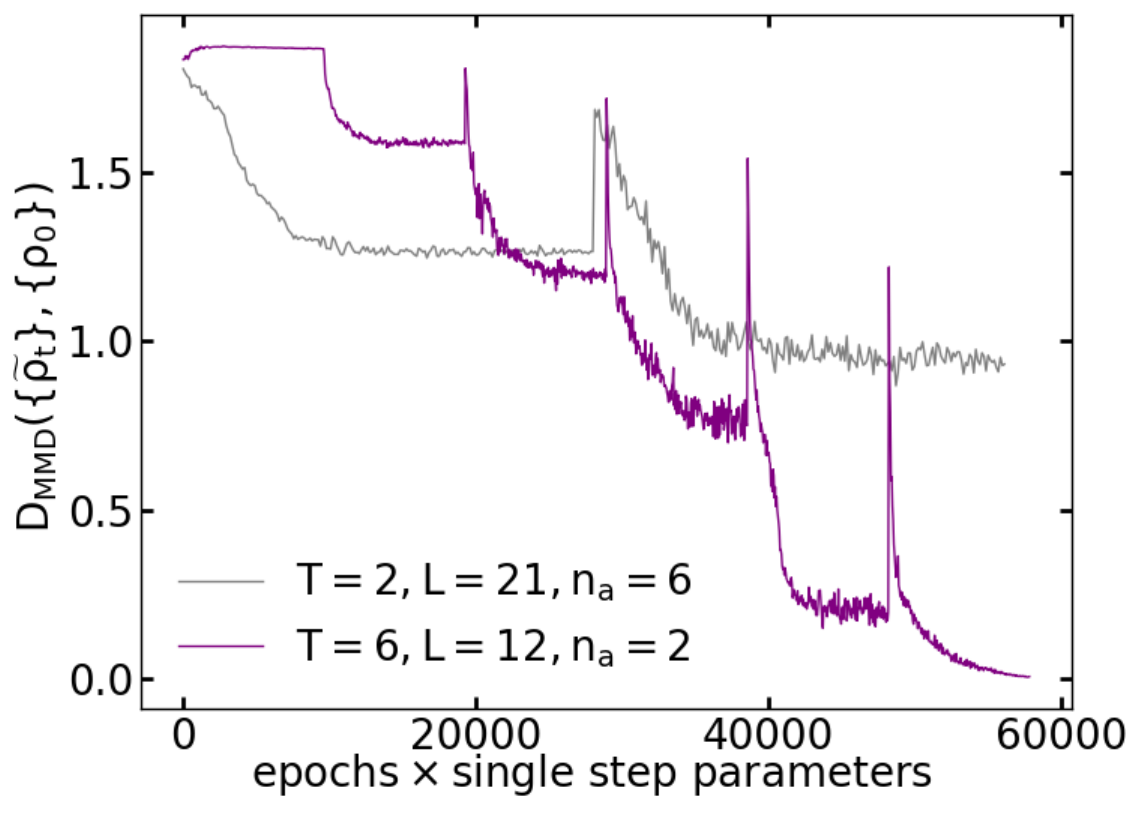}
\caption{\label{fig:diffusionComp}The MMD distance decay of models with different diffusion steps in the many-body phase learning task, computed between the training outputs $\{\tilde{\rho}_t\}$ and the samples from initial ensemble $\{\rho_0\}$. The purple line denotes the proposed model, which has a total of 864 parameters. The gray line represents the benchmark model with 840 parameters in total. Both models are trained with $n_{\rm train} = 50$ data. The periodic spikes are caused by the randomized initialization at each backward step.}
\end{figure}

Although the optimal number of diffusion steps may depend on the problem and hyperparameters, our results show that dividing the training process into multiple diffusion sequences is beneficial for efficiency and performance by reducing the actual qubits required for the backward VQC. The advantage of incorporating intermediate steps in MSQuDDPM is apparent in complex pattern-learning tasks. In Fig.~\ref{fig:diffusionComp}, we compare the performance of MSQuDDPM with different diffusion steps and the number of ancillary qubits in the many-body phase learning task. We benchmark a model against our proposed architecture with $n=4$ qubits, $T=6$ diffusion steps, $L=12$ layers in a single PQC, and $n_a = 2$ ancillary qubits. The configuration of the benchmark model is $n=4, T=2, L=21$, and $n_a = 6$, which employs more auxiliary qubits but fewer diffusion steps than the proposed model. To ensure a fair comparison, we maintain a similar total number of parameters across all models and use the total parameter updates, calculated as the product of iterations and the number of trainable parameters per step. As shown in Fig.~\ref{fig:diffusionComp}, the benchmark model that requires ten actual qubits presents training issues that preclude it from producing samples in the target distribution, resulting in a final MMD distance of 0.9325. In contrast, our proposed model with $T=6$ stages uses six actual qubits, considering that we can reset the auxiliary qubits after measurement and reuse them in the subsequent circuits. Our model better captures the target distribution than the benchmark, achieving the terminal MMD distance of 0.0061. Although both models utilize a total of 12 ancillary qubits, the benchmark model would not be robust enough for scaling to larger qubit systems, as the necessity of numerous qubits in a single circuit would make the model vulnerable to training problems and limit its capacity, especially when increasing the circuit depth \cite{larocca2024review}. While the relation between the model's capacity and hyperparameters, such as the number of ancillary qubits and diffusion steps, requires further investigation, increasing the diffusion steps can allow comparable or better performance in applying MSQuDDPM to the complex quantum state pattern generation.


\subsection{\label{sec:forwardSchedule} Noise scheduling in MSQuDDPM}

\begin{figure}
\includegraphics[width=0.5\textwidth]{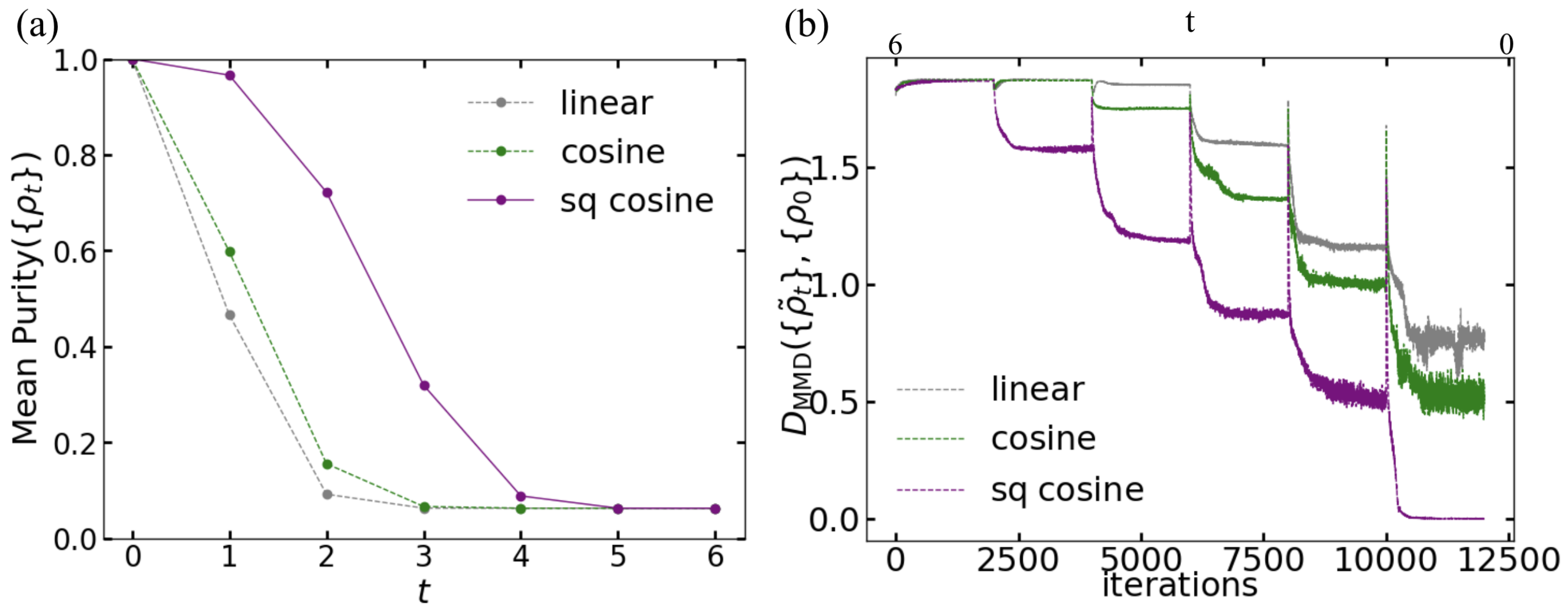}
\caption{\label{fig:forwardComp}Comparison of average purity decay for forward samples and MMD distance decay between the output ensemble of each training iteration and the initial state ensemble in many-body phase learning task with different forward schedulings. The gray, green, and purple lines in (a) represent the decay of the mean purity throughout the forward diffusion process of the 4-qubit many-body phase learning task with linear, cosine, and cosine square scheduling, respectively. The final mean purity for the $n=4$ qubits is $1/(2^4) = 0.0625$. In (b), the models utilize $n=4$ qubits, $n_a = 2$ ancillary qubits with $|00\rangle$ state, $T=6$, $L=12$, and MMD distance with superfidelity as a cost function. The gray, green, and purple lines indicate the MMD distance decay for linear, cosine, and cosine square scheduling, respectively. }
\end{figure}

In multi-qubit systems, the outputs of the MSQuDDPM's forward process with linear or cosine schedules tend to quickly lose their original characteristics, such as purity or magnetization. Fig.~\ref{fig:forwardComp}(a) illustrates the decay of average purity, defined as $\sum_{\rho \in \{\rho_t\}}\text{Tr}(\rho ^2) / |\{\rho_t\}|$, during the forward depolarizing process under different forward scheduling methods. The purities of linear and cosine schedules rapidly converge to the maximally mixed states by $t = 3$. Fast forgetting of the original form exacerbates the MSQuDDPM's performance, as it makes not only the later training rounds hard to converge but also the earlier training steps ineffective because the target state ensemble gathers close to the maximally mixed states instead of spreading out for better optimization. Thus, maintaining the original state is crucial when handling multi-qubit ensemble generation tasks with MSQuDDPM. To address these constraints, cosine-exponent scheduling, especially cosine square scheduling, is devised to better conserve the initial quantum states. In Fig.~\ref{fig:forwardComp}(a), we can observe that cosine square scheduling produces a smoother transition in average purity compared to other scheduling methods in the many-body phase learning task. 
\GK{A more detailed analysis of purity decay behavior under different noise scheduling methods is presented in Appendix~\ref{sec:noiseScheduleDetail}, where we extend the forward process simulations to other tasks. Our results indicate that while specific characteristics may vary depending on the task, the overall trend of purity decay can be universal across forward schedulings.}
In Fig.~\ref{fig:forwardComp}(b), we compare the MMD distance between the intermediate training results $\{\tilde{\rho}_t\}$ and the samples from the initial distribution $\{\rho_0\}$ across models using different noise scheduling methods in the same task. The results show significant differences in performance across the noise scheduling methods (see Table~\ref{tab:expresult} for numerical details). The model with cosine square scheduling successfully recovers the original ensemble, whereas the other models struggle to converge. 

\subsection{\label{sec:HaarAncilla}Haar random state as an ancillary qubit}

\begin{figure}
\includegraphics[width=0.45\textwidth]{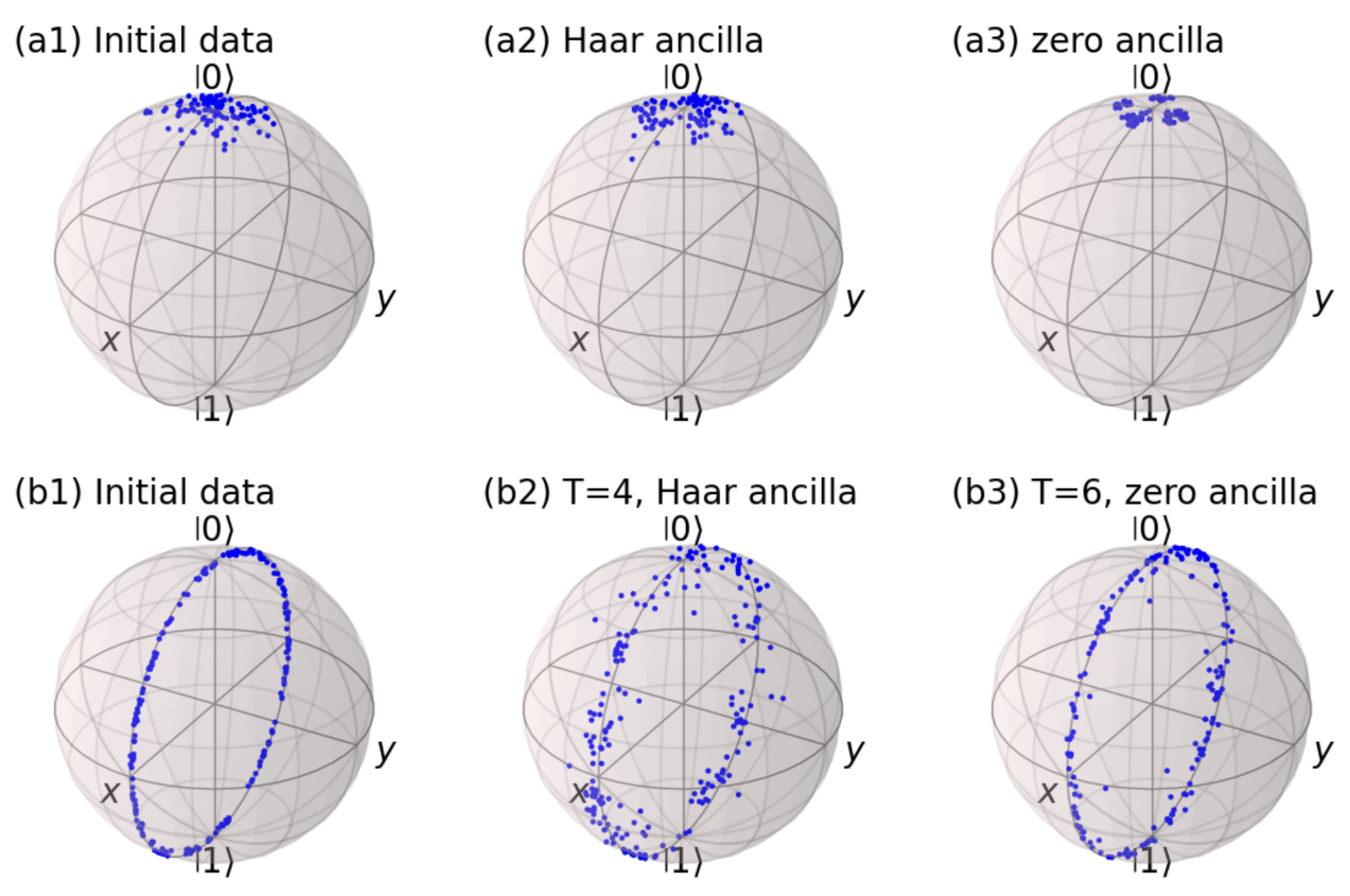}
\caption{\label{fig:HaarComp}MSQuDDPM test samples using different ancilla types for the clustered and circular state generation tasks ($n=1$). All models used $n_a=2$ auxiliary qubits. (a1) denotes the initial data for the cluster state generation task. (a2) and (a3) depict the backward test outputs with ancillary qubits initialized in a Haar random state ($|\phi_{\rm Haar}\rangle \otimes |0\rangle$) and the zero state ($|00\rangle$), respectively. (b1) - (b3) show the simulation results for the circular state generation task. (b1) represents the initial data, while (b2) and (b3) display the samples obtained using Haar random and zero-state ancilla qubits, respectively.}
\end{figure}

Although the initial ensemble for the backward PQC of MSQuDDPM, represented by the totally mixed state, can be viewed as having maximal noise, starting from identical states limits the model's ability to generate diverse samples in the early stages of the backward process. The number of possible outcomes at step $m$ (where $1 \leq m \leq T$) in the backward process, with $n_a$ auxiliary qubits initialized in $|0\rangle ^{\otimes n_a}$, is approximately $(2^{n_a})^{T-m+1} = 2^{n_a({T-m+1})}$, as measurements on the auxiliary qubits cause state branching. Consequently, the earlier rounds of backward learning (i.e., steps close to $T$) suffer from a capacity limit, as seen in the backward samples at $t=4$ in Fig.~\ref{fig:circular}(b). To address this issue, we leverage the Haar random state $|\phi_{\rm Haar}\rangle$ in auxiliary qubit initialization, expecting the backward PQC to manage the randomness induced by the entanglement between inputs and auxiliary qubits. We prepared $n_\text{train}$ 2-qubit ancillas ($n_a = 2$) with a single-qubit Haar random pure state ($|\phi_{\rm Haar}\rangle \otimes |0\rangle$). As described in Fig.~\ref{fig:HaarComp}, Haar ancillas introduce diversity into the model's outputs in Fig.~\ref{fig:HaarComp}(a1)-(a3) and allow the model to achieve performance comparable to that using zero-state ancillas but with fewer total parameters in Fig.~\ref{fig:HaarComp}(b1)-(b3) (see Table~\ref{tab:expresult} for numerical results). Therefore, using Haar ancilla can enhance the QuDDPMs' capabilities and serve as an additional strategy in auxiliary qubit utilization.

\section{\label{sec:conclusion}Discussions}

This work presents MSQuDDPM as a general and resource-efficient method for mixed quantum state ensemble generation. We demonstrate that this step-by-step denoising approach is beneficial to various quantum generative machine learning tasks. 
Through numerical experiments, we found that increasing the diffusion steps rather than the number of auxiliary qubits of MSQuDDPM results in better performance and robustness. 
We proposed cosine-exponent scheduling, especially cosine-square scheduling, which better preserves the original quantum states and is crucial for learning multi-qubit distributions. Moreover, Haar random states can serve as an additional ancilla option in VQC, as circuits with Haar ancillas are trainable, and the approach can further improve the model's capability.




We discuss several interesting future directions for MSQuDDPM. 
Further exploration of the architectures and hyperparameters in MSQuDDPM is an important direction for future research. Our numerical experiment on forward scheduling aligns with its classical counterpart, demonstrating that forward scheduling significantly influences the convergence of diffusion models \cite{chen2023importancenoiseschedulingdiffusion}. Hence, exploring the impact of cosine-exponent scheduling and its relation to other factors or approaches in MSQuDDPM would be a promising avenue for research. Also, the cost function of the proposed MSQuDDPM structure requires the estimation of superfidelity between two mixed-state ensembles, which introduces overhead in quantum experiments. 
Alternative and more efficiently measurable cost functions remain to be developed. 

In addition, the current architecture does not consider components such as positional encoding and the attention mechanism, which have been utilized in the DDPM \cite{ho2020denoisingdiffusionprobabilisticmodels} and latent diffusion model \cite{rombach2021highresolution}. Combining these elements with backward VQC through their quantum counterparts could enable even more complex quantum state generation. Furthermore, while Haar random states have shown their potential as auxiliary qubits in MSQuDDPM training, further investigation into their theoretical foundations and applications in other quantum optimization algorithms would be highly valuable. 

\GK{Another potential direction is to implement the proposed models on quantum devices and examine the scalability of the approach. The numerical simulations of MSQuDDPM are performed on classical computers, which require exponentially growing space to store and process mixed-state quantum data. Although the simulations demonstrate results up to four qubits due to resource limitations, we anticipate that MSQuDDPM can be applied to larger qubit systems when the individual VQC is designed to be sufficiently shallow, depending on the tasks and input qubits. Thus, deploying the models on quantum devices would provide insights into the method's scalability.}


\begin{acknowledgements}
QZ and BZ acknowledge support from NSF (CCF-2240641, OMA-2326746, 2350153), ONR N00014-23-1-2296 and DARPA (HR0011-24-9-0362, HR00112490453). This work was partially funded by an unrestricted gift from Google. QZ and BZ thank Xiaohui Chen and Peng Xu for discussions.
\end{acknowledgements}

\appendix

\section{\label{sec:circuitLayout}Circuit layout for forward and backward process }

\begin{figure}
\includegraphics[width=0.45\textwidth]{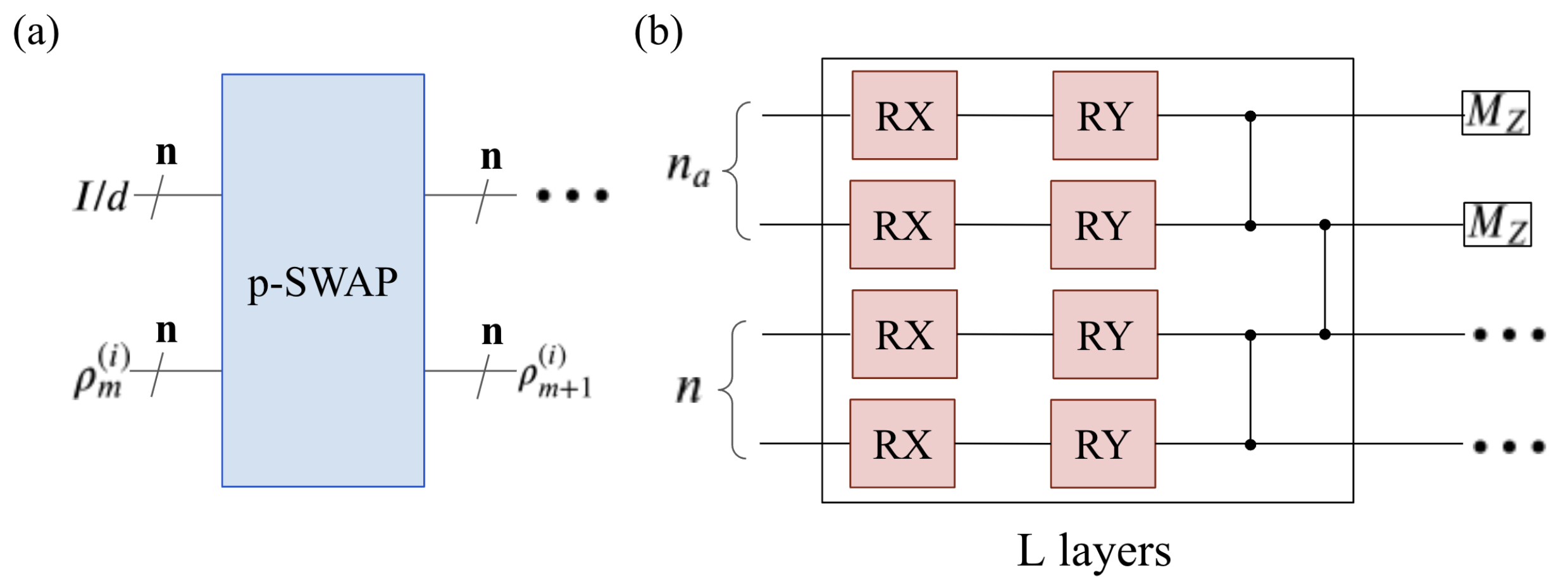}
\caption{\label{fig:circuitLayout}The detailed quantum circuits for the forward and backward processes in MSQuDDPM. In (a), one feasible implementation of a forward depolarizing channel for $n$ qubits using the p-SWAP gate is presented. (b) shows the backward PQC for $n = 2$ input qubits and $n_a = 2$ ancilla qubits. A single unitary layer consists of trainable RX and RY gates and imposes entanglement via CZ gates between neighboring qubits. After $L$ layers of unitaries, the model performs Z-axis projective measurements on all auxiliary qubits and collects the post-measurement density matrices.}
\end{figure}

In practice, we can construct the depolarizing channel of the forward process using a p-SWAP gate, as described in Fig.~\ref{fig:circuitLayout} (a). The p-SWAP gate performs a SWAP operation between two $n$-qubit inputs, $\rho_m^{(i)}$, and the fully mixed state $I/d$, with probability $q_{m+1}^{(i)}$, and applies the identity operator ($I$) with probability $1-q_{m+1}^{(i)}$.

Fig.~\ref{fig:circuitLayout} (b) depicts the detailed circuit diagram of the backward process. At each timestep $t$, the parameterized quantum circuit $\tilde{U}_t(\theta_t, \tilde{\rho}_t, |\tilde{\varphi}_t\rangle)$ incorporates $n_a$ ancillary qubits, $|\tilde{\varphi}_t\rangle$.  The tensor product of $|\tilde{\varphi}_t\rangle$ and $\rho_t^{(i)} \in \{\tilde{\rho}_t\} $ forms the complete input to a single PQC followed by applying $L$ layers of parameterized universal unitaries with hardware-efficient quantum ansatz \cite{kandala2017hardware, BingzhiQuDDPM} and entangling qubits. After the gates, Z-axis projective measurements are performed on the auxiliary qubits and the remaining density matrices form the output ensemble $\{\tilde{\rho}_{t-1}\}$, which is used as the input ensemble for the next timestep, ignoring the measurement outcomes.

\section{\label{sec:implDetail}Implementation details }

We implemented the MSQuDDPM using \texttt{Pytorch}, a widely-used machine learning library in \texttt{Python} \cite{pytorchNEURIPS2019}. For the quantum circuit implementation and automatic gradient calculation, we utilized the \texttt{TensorCircuit} library \cite{Zhang2023tensorcircuit}, and the Wasserstein loss calculation was based on the \texttt{POT} package \cite{flamary2021pot}. We visualized the outcomes in the Bloch sphere with \texttt{QuTip} \cite{Johansson_2012}. 


For optimization, we employed Adam optimizer \cite{kingma2014adam} with exponential learning rate decays since some proposals suggest that tuning the learning rate for Adam may be helpful for performance \cite{wilson2017marginal, loshchilov2017decoupled}. In the parameter initialization of the backward PQC, we considered normal and Xavier initialization methods \cite{pmlr-v9-glorot10a}, which include $n$ input qubits and $n_a$ ancillary qubits. For normal initialization, the parameters at training step $t$ are drawn from the standard Gaussian distribution, $\bm \theta_t \sim \mathcal{N}(0, 1)$. In Xavier initialization, weights are assigned as $\bm \theta_t \sim \mathcal{N}(0, \frac{1}{n+n_a})$,  using the total number of input and output qubits $n_{\rm in} = n_{\rm out} = n + n_a$ in $\mathcal{N}(0, \frac{2}{n_{\rm in}+n_{\rm out}})$. Note that even with Xavier initialization, we applied normal initialization to the ancillary qubits to facilitate random results in their projective measurements.

\begin{figure}
\includegraphics[width=0.48\textwidth]{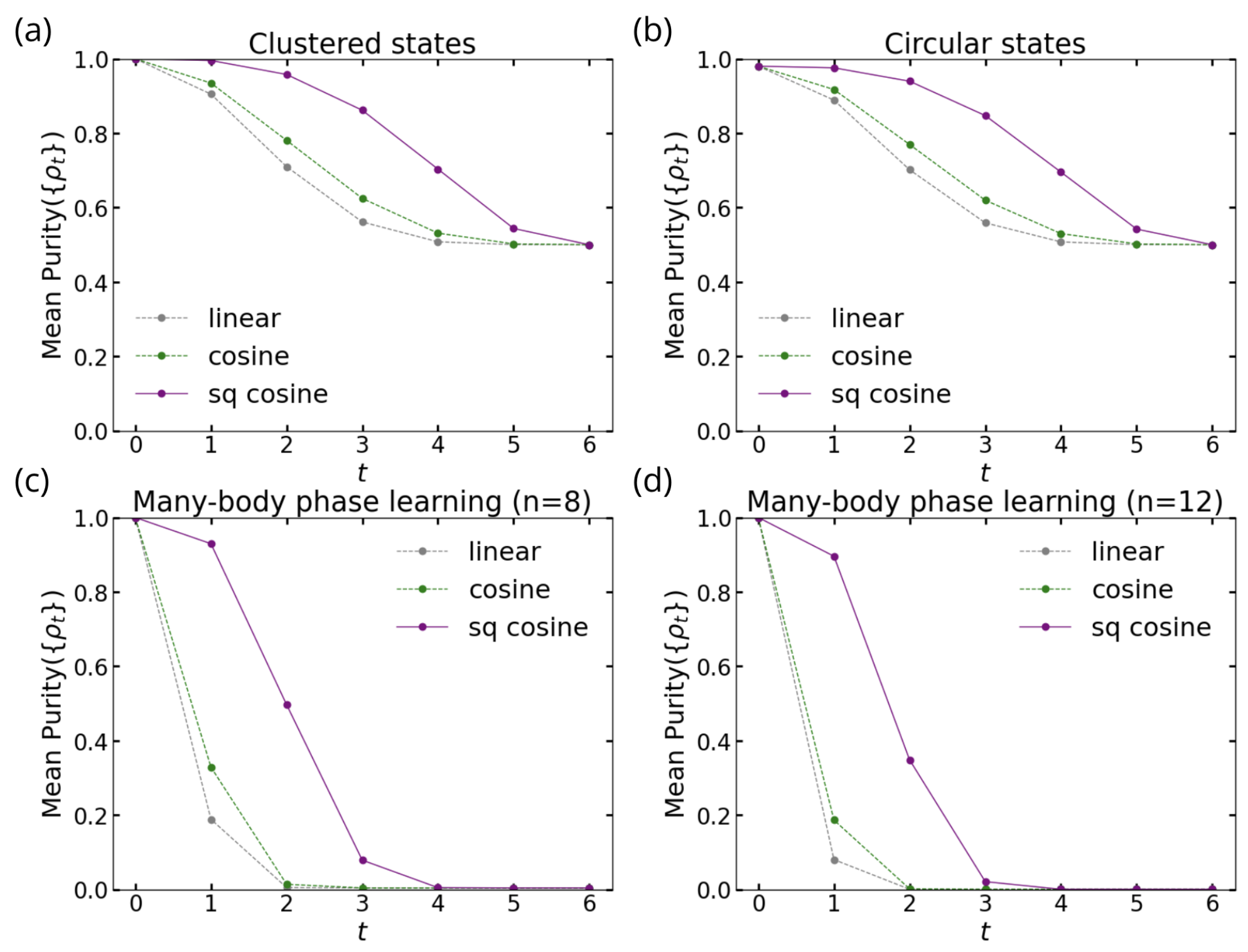}
\caption{\label{fig:forwardPurity} The mean purity decay of samples throughout the forward process of quantum state ensemble generation tasks. We used $n_\text{train}=n_\text{test}=100$ samples for each task, including (a) clustered states for $n=1$ qubit, (b) circular states for $n=1$ qubit, (c) many-body phase learning for $n=8$ qubits, and (d) many-body phase learning for $n=12$ qubits. In each plot, the gray, green, and purple lines represent the average purity decay under linear, cosine, and cosine square scheduling, respectively.}
\end{figure}

\section{\label{sec:noiseScheduleDetail}Additional details on noise scheduling in MSQuDDPM }

\GK{
In this section, we analyze the forward process with different noisy scheduling strategies considered in the main text. As shown in Fig.~\ref{fig:forwardPurity}, for different target state ensembles, the cosine square scheduling in Eq.~\eqref{eq:cosineExponent} enables a nontrivial states purity in more forward steps, which provides learnable information for the backward process. In contrast, the linear and cosine scheduling strategies undergo a sharp decay in the first few steps and make the states to be fully mixed in the late-stage, which in turn restricts the number of effective steps for learning in the backward process and thus results in limited performance as expected shown in Fig.~\ref{fig:forwardComp}. We expect the advantage from forward noise scheduling to be applicable to different learning tasks, and a detailed comparison of learning performance and a strategy for selecting an appropriate value of $k$ in Eq.~\eqref{eq:cosineExponent} remain as future work.}

\nocite{*}


%

\end{document}